\documentstyle[seceq,epsf]{ptptex}

\title{
Formation of Kaonic Atoms and Kaonic Nuclei in In-Flight ($K^-,p$) Reactions
}

\author{
Junko {\sc Yamagata}$^{1}$, Hideko {\sc Nagahiro}$^{2}$, 
Yuko {\sc Okumura}$^{1,}$\footnote{ Present address:  Ritto High School, Ritto, Shiga, Japan}  and Satoru {\sc Hirenzaki}$^{1}$
}

\inst{
$^{1}$Department of Physics, Nara Women's University, Nara 630-8506, Japan\\
$^{2}$Research Center for Nuclear Physics(RCNP), Osaka University, Ibaraki, Osaka 567-0047, Japan\\

}

\recdate{
March 11, 2005
}

\abst{
We theoretically study kaonic atom and kaonic nucleus formation in in-flight ($K^-, p$) reactions for C,
O, Si and Ca targets. Deeply bound kaonic atoms were previously predicted to 
exist as quasi-stable  states, and it is expected that they can be observed using certain 
well-suited experimental methods.  Kaonic nuclear states have also been predicted to 
exist with large decay widths.  We evaluate the formation cross sections 
of kaonic atoms and kaonic nuclei using an effective number approach.  
We show that indications of kaonic bound states can be observed 
in the outgoing proton energy spectra.  
}

\begin{document}

\maketitle

\section{Introduction}
\label{sec:intro}
Kaonic atoms and kaonic nuclei 
carry important information concerning the $K^-$-nucleon interaction 
in nuclear medium.  This information is very important to determine the 
constraints on kaon condensation in high density matter. The properties 
of kaons in nuclei are strongly influenced by the change undergone by 
$\Lambda(1405)$ in nuclear medium, because $\Lambda(1405)$ is a resonance 
state just below the kaon-nucleon threshold.  In fact, there are studies 
of kaonic atoms carried out by modifying the properties of $\Lambda(1405)$ in nuclear medium.
\cite{alberg76,wei,miz} These works reproduce the properties of kaonic atoms very well, 
which come out to be as good as the 
phenomenological study of Batty.\cite{bat}  

Recently, there have been significant developments in the description
of hadron properties in terms of the $SU(3)$ chiral Lagrangian.
The unitarization of the chiral Lagrangian allows the interpretation of the
$\Lambda(1405)$ resonance state as a baryon-meson coupled 
system.\cite{kai,ose}  Subsequently, the properties of $\Lambda(1405)$
in nuclear medium 
using the $SU(3)$ chiral unitary model were also investigated by
Waas et al.,\cite{waa} \ Lutz,\cite{lut} Ramos and Oset,\cite{ram} and Ciepl$\acute{\rm y}$ et al.\cite{ciep01} \ 
All of these works considered the Pauli effect on the 
intermediate nucleons. In addition, in Ref.~\citen{lut}, the self-energy of the 
kaon in the intermediate states is considered, and in Ref.~\citen{ram}, the 
self-energies of the pions and baryons are also taken into account.  These 
approaches lead to a kaon self-energy in nuclear medium that can be 
tested with kaonic atoms and kaonic nuclei. There are also $\bar{K}$ potential studies based on meson-exchange J$\ddot{\rm u}$lich $\bar{K}N$ interaction.\cite{tolo01,tolo02}

In a previous work,\cite{hirenzaki00} \ we adopted the scattering amplitude in nuclear 
medium calculated by Ramos and Oset\cite{ram} for studies of
kaonic atoms, and demonstrated the ability to
reproduce the existing kaonic atom data as accurately as the optical potential
studied by Batty.\cite{bat} \ We then calculated the deeply bound kaonic
atoms for $^{16}$O and $^{40}$Ca, which have narrow widths 
and are believed to be observable with well-suited experimental methods.\cite{16.5_Friedman99,Friedman99} \ 
We also obtained
very deep kaonic nuclear states, which have large decay widths, 
of the order of several tens of MeV. The $(K^-, \gamma)$ reaction 
was studied for the formation of the deeply bound kaonic atomic
states,\cite{hirenzaki00} \ which could 
not be observed with kaonic X-ray spectroscopy, using the formulation
developed in Ref.~\citen{nie} for the formation of 
deeply bound
pionic atoms with the  $(\pi^-, \gamma)$ reaction.

Another very important development in recent years is that in the study 
of kaonic nuclear states, which are kaon-nucleus bound systems determined mainly 
by the strong interaction. Experimental studies of the kaonic nuclear 
states using in-flight ($K,N$) reactions were proposed and performed by Kishimoto and his collaborators.\cite{Kishimoto99,Kishimoto03} \ Experiments employing stopped ($K,N$) reactions were carried out 
by Iwasaki, T. Suzuki and their collaborators and reported in Refs.~\citen{Iwasaki03} and \citen{Suzuki04}.\ 
In these experiments, they found some indications of the existence of kaonic nuclear states.  
There are also theoretical studies of the structure and formation of kaonic nuclear states related to 
these experimental activities.\cite{Akaishi02} \ It should be noted that these theoretical studies 
predict the possible existence of ultra-high density states in kaonic nuclear systems. \cite{Akaishi02,dote04}

In this paper, we study in-flight ($K^-, p$) reactions systematically with regard to their role in populating 
deeply bound kaonic states and the observation of their properties in experiments. 
We have found the usefulness of 
direct reactions in the formation of deeply bound pionic atoms 
using the ($d,^3$He) reactions.\cite{tok,hirenzaki91,gilg00,itahashi00} \ 
However, in the present case, $K^+$ must be produced in 
addition in this ($d,^3$He) reaction, 
and there would be a large momentum mismatch.  For this reason, 
the $(K^-, \gamma)$ reaction was considered first in Ref.~\citen{hirenzaki00}. \ 
Here we theoretically study another reaction, ($K^-, p$), and present systematic results that elucidate 
the experimental feasibility of the reaction.  The ($K^-, p$) reaction 
was proposed in Refs.~\citen{Friedman99} and \citen{Kishimoto99}. \ However, 
realistic spectra have not yet been calculated. We calculate the 
spectra theoretically using the approach of Ref.~\citen{hirenzaki91} 
for the deeply bound pionic atom formation reaction.  
We believe that this theoretical evaluation will be interesting and important for 
studies of kaon properties in nuclear medium.

In $\S$\ref{sec:structure}, we describe the theoretical model of the structure of kaon-nucleus 
bound systems and present the numerical results. The theoretical formalism and numerical results for the 
($K^-,p$) reactions are discussed in $\S$\ref{sec:formation}. We give summary in $\S$\ref{sec:conclusion}.
\section{Structure of kaonic atoms and kaonic nuclei}
\label{sec:structure}

\subsection{Formalism}
\label{S_Form}
We study the properties of kaonic bound systems by solving the Klein-Gordon equation

\begin{equation}
[-{\bf {\nabla}}^2+\mu^2+2\mu V_{\rm opt}(r)]\phi(\mbox{\boldmath $r$})=[E-V_{\rm coul}(r)]^2 \phi(\mbox{\boldmath $r$})~.
\label{KGeq}
\end{equation}
\noindent
Here, $\mu$ is the kaon-nucleus reduced mass and $V_{\rm coul}(r)$ is the Coulomb potential with a 
finite nuclear size:
\begin{equation}
V_{\rm coul}(r)=-e^2 \int \frac{\rho_p(r')}{|\mbox{\boldmath $r$-$r'$}|}d^3 r'~,
\label{V_coul}
\end{equation}

\noindent
where $\rho_p(r)$ is the proton density distribution. We employ the empirical Woods-Saxon form 
for the density and keep the shapes of the neutron and proton density distributions fixed as

\begin{equation}
\rho (r) = \rho_n(r)+\rho_p(r) = \frac{\rho_{\rm 0}}{1+\exp[(r-R)/a]}~,
\label{rho}
\end{equation}

\noindent
where we use $R=1.18A^{1/3}-0.48~[{\rm fm}]$ and $a=0.5~[{\rm fm}]$ with $A$, the nuclear mass number. It is noticed that the point nucleon density distributions are 
deduced from $\rho$ in Eq. (\ref{rho}) by using the same prescription described in Sect. 4 in Ref.~\citen{nieves93} and are used to evaluate the kaon-nucleus optical potential.

The kaon-nucleus optical potential $V_{\rm opt}$ is given by

\begin{equation}
2\mu V_{\rm opt} (r) = -4 \pi \eta a_{\rm eff}(\rho)\rho(r) , 
\label{V_opt}
\end{equation}
\noindent
where $a_{\rm eff}$($\rho$) is a density dependent effective scattering length and $\eta=1+m_K/
M_N$.
In this paper, we use two kinds of effective scattering lengths, that obtained with 
the chiral unitary approach\cite{ram} and that obtained with a phenomenological fit.\cite{batty97}~\ Here, we do not introduce any energy dependence for the 
effective scattering lengths, and we use the scattering lengths at the $KN$ 
threshold energy. 
The effective scattering length $a_{\rm eff}$ of the chiral unitary approach is described in Ref.~\citen{hirenzaki00} \ in detail. It 
is defined by the kaon self-energy in nuclear matter, with the local density approximation. The form of 
$a_{\rm eff}$ obtained in a phenomenological fit is one of the results reported in Ref.~\citen{batty97}, \ and it 
is parameterized as
\begin{equation}
a_{\rm eff}(\rho)=(-0.15+0.62i)+(1.66-0.04i)(\rho/\rho_{\rm 0})^{0.24} [{\rm fm}] .
\label{batty_a}
\end{equation}
\noindent
The reason we consider these two potentials is that they provide 
equivalently good descriptions of the observed kaonic atom data, even though they have very 
different potential depths, as we will see in next subsection. 
Thus, it should be extremely interesting to compare the results obtained with these potentials 
in the ($K^-,p$) reaction spectra, including the kaonic nuclear region.

We solve the Klein-Gordon equation numerically, following the method of Oset and Salcedo.\cite{oset85} \ 
The application of this method to pionic atom studies are reported in detail in Ref.~\citen{nieves93}.

\subsection{Numerical results}
\label{S_results}

\begin{figure}[htbp]
\epsfysize=5cm
\centerline{\epsfbox{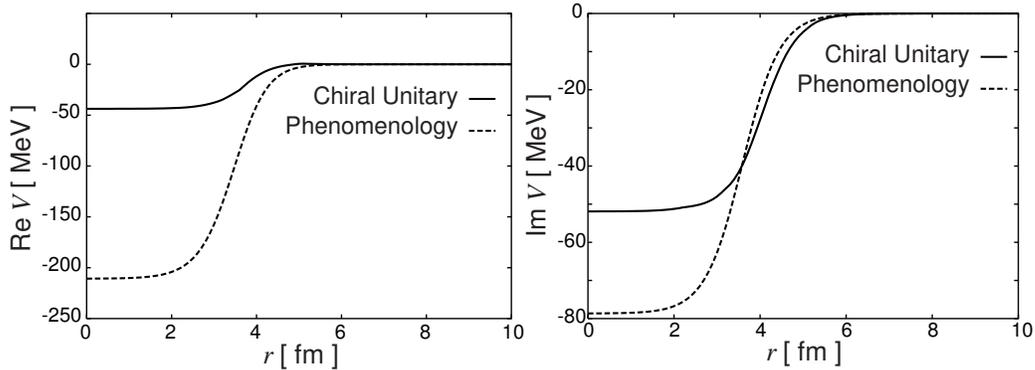}}
\caption{The kaon-nucleus optical potential for $^{39}$K as a function of the radial coordinate $r$.
The left and right panels show the real and imaginary part, respectively. 
The solid line indicates the potential strength of the chiral unitary approach and the dashed line 
of the phenomenological fit.}
\label{fig_V_opt}
\end{figure}

We show the kaon nucleus potential for the $^{39}$K case in Fig.~\ref{fig_V_opt} as an example. 
Because the real part of $a_{\rm eff}$($\rho$) changes sign at a certain nuclear density in 
both the chiral unitary and phenomenological models, the kaon nucleus optical potential is 
attractive, while keeping the repulsive sign for the kaon-nucleon scattering length in free space.

The real part of the scattering length for the phenomenological fit depends on the density much 
more strongly than the results of the chiral unitary model and yields Re $a_{\rm eff}$($\rho_0)=1.51 [{\rm fm}]$.
Hence, as we can see in Fig.~\ref{fig_V_opt}, \ the depths of the real 
optical potentials of these models differ significantly.
 On the other hand, the density 
dependence of the imaginary part of the phenomenological 
scattering length is rather flat, and its strength is similar to that of 
the chiral unitary model.

\begin{figure}[htbp]
\epsfxsize=10cm
\centerline{\epsfbox{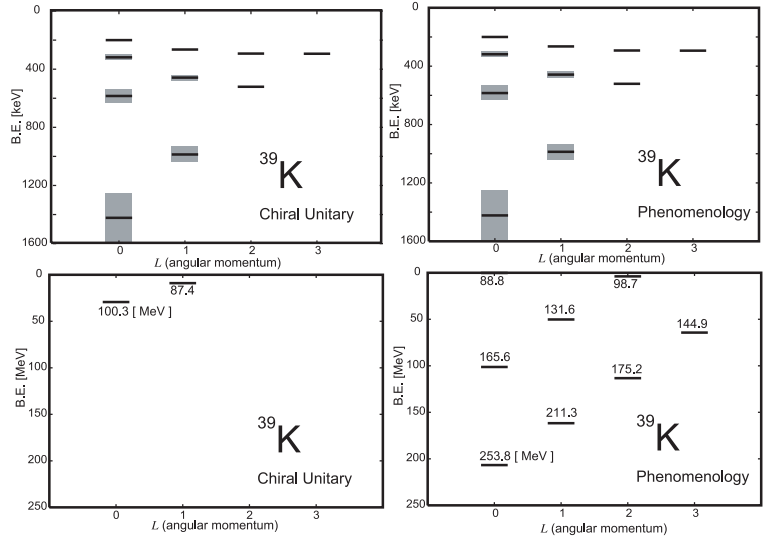}}
\caption{(Upper panel) Energy levels of kaonic atoms of $^{39}$K obtained with the optical 
potentials of the chiral unitary model (left) and the phenomenological fit (right). 
The hatched areas indicate the level widths. 
(Lower panel) Energy levels of kaonic nuclear states of $^{39}$K obtained with the optical 
potentials of the chiral unitary model (left) and the phenomenological fit (right). 
The level width is indicated by the number appearing at each level, in units of MeV.}
\label{fig:39K_Energy}
\end{figure}

The calculated energy levels for the atomic states and nuclear states 
in $^{39}$K are shown in Fig.~\ref{fig:39K_Energy}, where the results of the chiral 
model and the phenomenological model, Eq.~(\ref{batty_a}), are compared. 
We see that the results obtained with the two potentials are very similar for the atomic states. 
We find that the deep atomic states, such as atomic 1$s$ in $^{39}$K (still unobserved), 
appear with narrower widths than the separation between levels and are predicted 
to be quasi-stable states. Similar results are reported in previous works.
\cite{hirenzaki00,16.5_Friedman99} \ Because several model potentials predict 
the existence of quasi-stable deep atomic states, it would be interesting to 
observe the states experimentally. On the other hand, the predicted binding 
energies and widths are very stable and almost identical for all of the potential models 
considered here. 
Hence, it is very difficult to distinguish the theoretical potentials from only 
the observation of atomic levels.

In the lower panels of Fig.~\ref{fig:39K_Energy}, we also show the energy levels 
of the deep nuclear kaonic states of $^{39}$K using the chiral unitary model 
potential and the phenomenological model potential. The deep nuclear states 
are represented by the solid bars with numbers, which indicate their widths in units of MeV. 
These nuclear states have extremely large widths in all cases and would not be observed 
as peak structures in experiments if they indeed do have such large widths. We should, 
however, mention here that the level structures of these potential models differ 
significantly. In the chiral unitary potential, only two nuclear 
states are predicted, while eight states are predicted with the phenomenological 
model. This difference presents the opportunity to distinguish the theoretical potentials in 
observations of kaonic nuclear states. 

\begin{figure}[htpd]
\epsfxsize=6cm
\centerline{\epsfbox{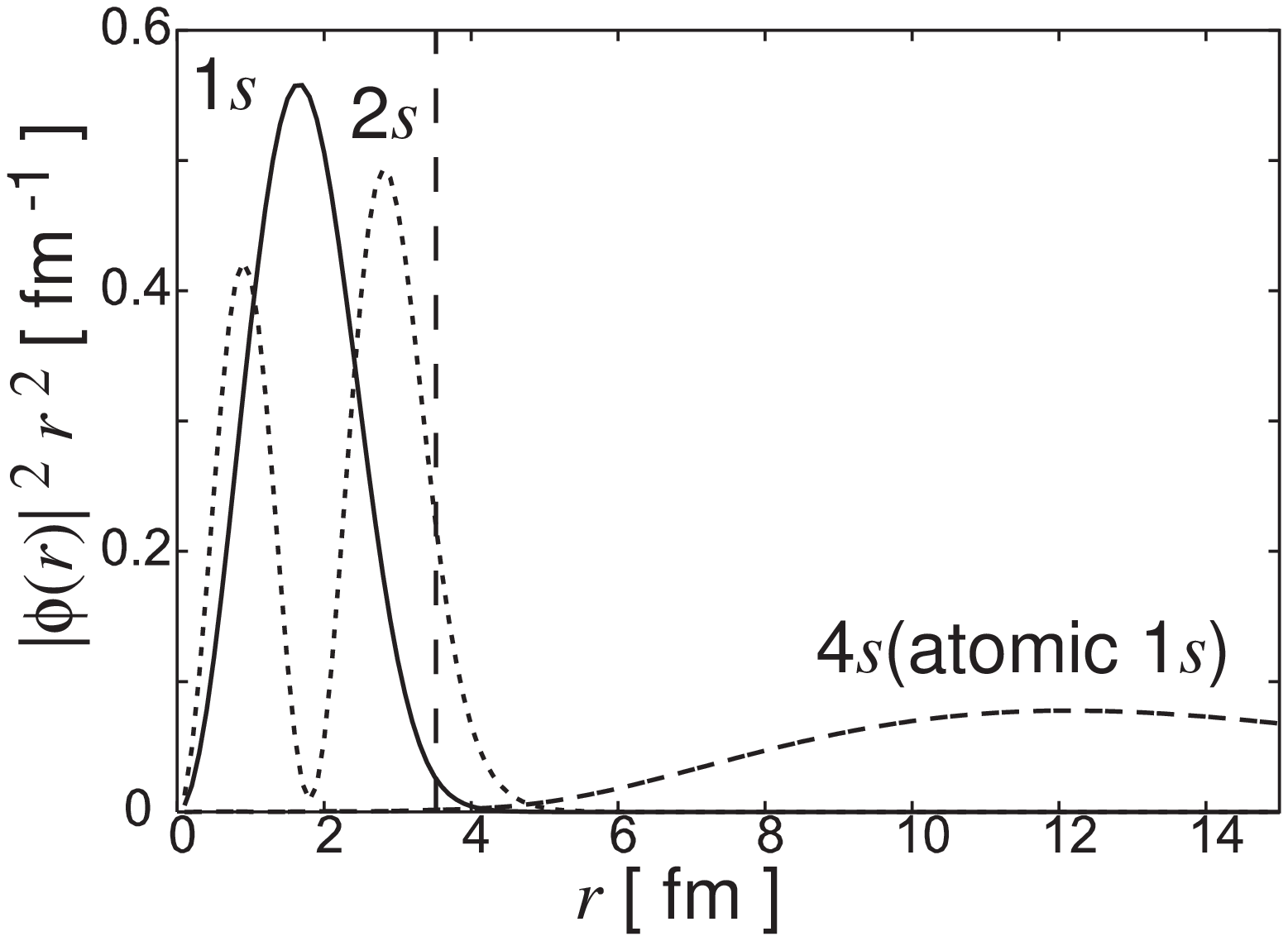}}
\caption{The kaonic bound state density distributions 
$|r\phi(r)|^2$ in coordinate space for $^{39}$K obtained with 
the phenomenological optical potential. The 
solid and dotted curves indicate the distributions of the 1$s$ and 2$s$ states. The dashed 
curve represents the density of the 4$s$ state and is regarded as a kaonic atom 1$s$ state. The 
half-density radius of $^{39}$K is also shown.}
\label{fig:39K_wf}
\end{figure}

Calculated density distributions of nuclear 1$s$ and 2$s$ and atomic 1$s$ kaonic 
states in $^{39}$K are shown in Fig.~\ref{fig:39K_wf} for the case of the 
phenomenological optical potential. It is seen that the wavefunctions of the deep nuclear kaonic 
states remain almost entirely inside the nuclear radius, which is about 3.5 fm for 
$^{39}$K. Hence, the widths become extremely large, of the order of 100 MeV. 
The wavefunctions of the atomic states are pushed outward by the imaginary part of 
the strong interaction. It should be noted that the atomic 1$s$ state 
corresponds to the 4-th $s$ state in the solutions of the Klein-Gordon 
(KG) equation, Eq.~(\ref{KGeq}). We divided the series of KG solutions into two categories, 
'atomic states' and 'nuclear states', since the properties of these 
states are very different, and there are no ambiguities in this 
classification, as can be seen in Figs. \ref{fig:39K_Energy} and \ref{fig:39K_wf}.

We have also calculated the kaon-nucleus binding energies and widths for both 
atomic and nuclear states in other nuclei. 
The obtained results are compiled in Tables~\ref{tab:ph} and \ref{tab:chi} 
for the phenomenological optical potential and for the chiral unitary potential 
cases, respectively. 
We selected $^{11}$B,~$^{15}$N,~$^{27}$Al and $^{39}$K nuclei, which appear 
in the final states of the formation reactions for $^{12}$C,~$^{16}$O,~$^{28}$Si 
and $^{40}$Ca targets, as described in the next section. 
In all cases, we found kaonic atom states and kaonic nuclear states. The 
results for the atomic states are similar for the two potentials and 
are known to reproduce existing data reasonably well. On the other hand, 
certain differences are found in the energy spectra of kaonic nuclear states, 
as expected, and they should be investigated experimentally.

\begin{table}[htbp]
\begin{center}
\caption{Calculated binding energies and widths of kaon-$^{11}$B, -$^{15}$N, 
-$^{27}$Al and -$^{39}$K systems with the phenomenological optical potential 
in units of MeV for kaonic nuclear states and in units of keV for kaonic atom states.}
\begin{tabular}{c|cc|cc|cc|cc}
\hline
\hline
&&&&&&&&\\
Nuclear State& \multicolumn{2}{|c|}{$^{11}$B}&\multicolumn{2}{|c|}{$^{15}$N}
&\multicolumn{2}{|c|}{$^{27}$Al}&\multicolumn{2}{|c}{$^{39}$K}\\ 
(MeV)&B.E.& $\Gamma$&B.E.& $\Gamma$&B.E.& $\Gamma$&B.E.& $\Gamma$ \\ 
\hline
1$s$ & 132.5 & 183.0 & 155.7 & 205.5&190.8&239.5&206.7&253.8\\
2$s$&-&-&19.0&96.2&69.8&142.3&101.3&165.6\\
3$s$&-&-&-&-&-&-&2.1$\times$10$^{-1}$&88.8\\
2$p$&58.4&127.0&86.9&151.1&136.3&191.2&161.7&211.3\\
3$p$&-&-&-&-&16.5&103.3&50.2&131.6\\
3$d$&-&-&22.9&108.3&80.4&152.2&113.3&175.2\\
4$d$&-&-&-&-&-&-&3.9&98.7\\
4$f$&-&-&-&-&26.2&119.5&64.3&144.9\\
\hline
Atomic State&&&&&&&&\\
(keV)&&&&&&&&\\ 
\hline
1$s$&192.4&40.6&338.1&68.3&844.9&243.1&1408.0&355.4\\
2$s$&60.5&7.3&111.8&13.3&319.0&58.3&580.6&97.6\\
3$s$&29.2&2.5&55.1&4.6&166.8&22.3&316.9&39.9\\
4$s$&17.1&1.1&21.7&1.1&102.3&10.8&199.4&20.0\\
2$p$&78.5&6.2$\times$10$^{-1}$&154.6&4.0&507.5&37.4&988.7&132.7\\
3$p$&34.9&2.2$\times$10$^{-1}$&68.8&1.4&229.4&12.7&458.4&45.8\\
4$p$&19.6&9.6$\times$10$^{-2}$&38.7&6.1$\times$10$^{-1}$&130.5&5.7&265.1&20.8\\
3$d$&34.9&2.3$\times$10$^{-4}$&69.3&2.9$\times$10$^{-3}$&243.0&4.2$\times$10$^{-1}$&520.9&5.0\\
4$d$&19.6&1.4$\times$10$^{-4}$&39.0&1.7$\times$10$^{-3}$&136.6&2.5$\times$10$^{-1}$&292.6&2.9\\
4$f$&19.6&1.0$\times$10$^{-8}$&38.9&4.8$\times$10$^{-7}$&136.5&3.2$\times$10$^{-4}$&293.7&1.7$\times$10$^{-2}$\\
\hline
\end{tabular}
\label{tab:ph}
\end{center}
\end{table}

\begin{table}[htpb]
\begin{center}
\caption{Calculated binding energies and widths of kaon-$^{11}$B, -$^{15}$N, 
-$^{27}$Al and -$^{39}$K systems with the optical potential of the chiral unitary model, 
in units of MeV for kaonic nuclear states and in units of keV for kaonic atom states.}
\begin{tabular}{c|cc|cc|cc|cc}
\hline
\hline
&&&&&&&&\\
Nuclear State& \multicolumn{2}{|c|}{$^{11}$B}&\multicolumn{2}{|c|}{$^{15}$N}
&\multicolumn{2}{|c|}{$^{27}$Al}&\multicolumn{2}{|c}{$^{39}$K}\\ 
(MeV)&B.E.& $\Gamma$&B.E.& $\Gamma$&B.E.& $\Gamma$&B.E.& $\Gamma$ \\ 
\hline
1$s$&4.6&81.6&11.0&87.9&22.5&96.6&29.3&100.3\\
2$p$&-&-&-&-&-1.1&79.5&9.0&87.4\\
\hline
Atomic State&&&&&&&&\\
(keV)&&&&&&&&\\
\hline
1$s$&197.5&35.0&340.5&67.8&852.1&199.1&1422.7&337.3\\
2$s$&61.3&6.2&112.3&13.2&320.4&47.6&584.3&92.5\\
3$s$&29.5&2.1&55.3&1.6&167.2&18.2&318.3&37.8\\
4$s$&17.3&9.3$\times$10$^{-1}$&21.7&1.1&102.5&8.8&200.1&19.0\\
2$p$&78.4&6.3$\times$10$^{-1}$&154.4&3.1&507.3&34.8&986.8&109.3\\
3$p$&34.8&2.2$\times$10$^{-1}$&68.7&1.1&229.3&11.8&457.7&37.7\\
4$p$&19.6&9.7$\times$10$^{-2}$&38.7&4.8$\times$10$^{-1}$&130.5&5.3&264.8&17.1\\
3$d$&34.9&1.6$\times$10$^{-4}$&69.3&2.6$\times$10$^{-3}$&243.0&3.4$\times$10$^{-1}$&520.8&4.5\\
4$d$&19.6&9.5$\times$10$^{-5}$&39.0&1.6$\times$10$^{-3}$&136.6&2.0$\times$10$^{-1}$&292.6&2.6\\
4$f$&19.6&1.0$\times$10$^{-8}$&38.9&3.7$\times$10$^{-7}$&136.5&2.8$\times$10$^{-4}$&293.7&1.5$\times$10$^{-2}$\\
\hline
\end{tabular}
\label{tab:chi}
\end{center}
\end{table}

We should mention here that the kaonic nuclear 2$p$ state in $^{27}$Al described in 
Table \ref{tab:chi} provides a negative value for the binding energy. 
This state, however, is interpreted as a bound state, since the sign of the corresponding eigenenergy 
in the non-relativistic Schr$\ddot{\rm o}$dinger equation is opposite 
to that of the Klein-Gordon solution, due to the large widths of the nuclear states, as shown below. 
The binding energies $B_{\rm KG}$ and widths $\Gamma_{\rm KG}$ of the 
Klein-Gordon equation, which are tabulated in Tables \ref{tab:ph} and 
\ref{tab:chi}, are defined as $E=(\mu - B_{\rm KG})-\frac{i}{2}\Gamma_{\rm KG}$ 
by the eigenenergy $E$ in Eq.~(\ref{KGeq}). The non-relativistic binding energy 
$B_{\rm S}$ and width $\Gamma_{\rm S}$ of the Schr$\ddot{\rm o}$dinger 
equation are related to $B_{\rm KG}$ and $\Gamma_{\rm KG}$ as

\begin{equation}
B_{\rm S}=B_{\rm KG}-\frac{B_{\rm KG}^2}{2\mu}+\frac{\Gamma_{\rm KG}^2}{8\mu} ,
\label{eq:E_s-KG}
\end{equation}

\begin{equation}
\Gamma_{\rm S}=\Gamma_{\rm KG}-\frac{B_{\rm KG}}{\mu}\Gamma_{\rm KG}.
\label{eq:G_s-KG}
\end{equation}
\noindent
Thus, in the case of the kaonic 2$p$ nuclear state in $^{27}$Al described in Table 
\ref{tab:chi}, the non-relativistic binding energies and widths 
are $B_{\rm S}\sim 0.5$MeV 
and $\Gamma_{\rm S}\sim \Gamma_{\rm KG}$, which indicate that the state is 
bound. It should be noted that the asymptotic behavior of the wavefunction 
is determined by $B_S$.

\section{Kaonic atoms and kaonic nuclei formation in ($K^-,p$) reactions}
\label{sec:formation}

\subsection{Formalism}
We adopt the theoretical model presented in Ref.~\citen{hirenzaki91} 
to calculate the formation cross sections of the kaonic atoms and kaonic nuclei 
in the ($K^-, p$) reaction. 
  In this model, the emitted proton energy 
spectra can be written as 

\begin{equation}
\frac{d^2\sigma}{dE_ p d\Omega_ p} = 
\left( \frac{d \sigma}{d \Omega} \right)^{\rm lab}_{K^- p \rightarrow p K^-} 
\sum_f \frac{ \Gamma_ K}{2 \pi} 
\frac{1}{\Delta E^2 + \Gamma_ K^2/4} N_{\rm eff} , 
\label{eqn:cross-section}
\end{equation}
\noindent
where the sum is over all (kaon-particle) $\otimes$ (proton-hole) configurations in the 
final kaonic bound states. 
The differential cross section for the elementary process of the reaction 
$K^- + p $ $\rightarrow$ $p + K^-$ in the laboratory frame, 
$( d \sigma/d \Omega)^{\rm lab}_{K^- p \rightarrow p K^-} $ 
is evaluated using the $K^-p$ total elastic cross section data in 
Ref.~\citen{PDG2004} \ by assuming a flat angular distribution in the 
center-of-mass frame at each energy. 
The resonance peak energy is determined by $\Delta E$ appearing 
in Eq.~(\ref{eqn:cross-section}), 
which is defined as 

\begin{equation}
\Delta E = T_ p - (T_ K - S_ p (j_ p^{-1}) + B_ K) , 
\label{eq:DE}
\end{equation} 
\noindent 
where $T_ K$ is the incident kaon kinetic energy, $T_ p$ the emitted 
proton kinetic energy, and $B_ K$ the kaon binding energy in the final state. 
The proton separation energy $S_ p$ from each single particle 
level listed in Table~\ref{tb:sp} is obtained from the data in 
Refs.~\citen{Belo85,yosoi04,ajz91,mougey76,amaldi64,nakamura74}. We use the data in 
Ref.~\citen{TOI96} for the separation energies of the proton-hole 
levels corresponding to the ground states of the daughter nuclei. The widths of the hole states $\Gamma_ p$ are 
also listed in Table~\ref{tb:sp}. These were obtained from the same data 
sets by assuming the widths of the ground states of the daughter nuclei to 
be zero because of their stabilities.

\begin{center}  
\begin{table}[hpbt]
\begin{center}
\caption{One proton separation energies $S_ p$ and widths $\Gamma_ p$ 
of the hole states 
of $^{12}$C, $^{16}$O, $^{28}$Si and $^{40}$Ca 
deduced from the data given in Ref.~\protect \citen{Belo85} \ for $^{12}$C, those given 
in Refs.~\protect \citen{yosoi04} and~\protect \citen{ajz91} \ for $^{16}$O, 
those given in Refs.~\protect \citen{mougey76} and~\protect \citen{amaldi64} \ for $^{28}$Si, 
and those given in Ref~\protect \citen{nakamura74} \ for $^{40}$Ca. 
The separation energies corresponding to the ground states of the daughter 
nuclei are taken from Ref.~\protect \citen{TOI96}. 
The widths $\Gamma_ p$ indicate FWHM of the Lorentz distribution for 
$^{16}$O and of Gaussian distributions for other nuclei. 
The widths of the ground states of the daughter nuclei are fixed 
to zero, because of their stabilities. 
For the 1$p$ states in $^{28}$Si and $^{40}$Ca, two levels, 1$p_{1/2}$ and 1$p_{3/2}$, 
have not been observed separately, and therefore $S_ p$ and $\Gamma_ p$ 
are set to the same values for both levels.
}
\vspace{3mm}
\begin{tabular}{|c|cc|cc|cc|cc|} 
\hline
&&&&&&&&\\
   single particle   & \multicolumn{2}{|c|}{$^{12}$C}&\multicolumn{2}{|c|}{$^{16}$O}&
\multicolumn{2}{|c|}{$^{28}$Si}&\multicolumn{2}{|c|}{$^{40}$Ca}\\
 states [MeV]& $S_ p$ & $\Gamma_ p$ & $S_ p$ & $\Gamma_ p$
&$S_ p$ & $\Gamma_ p$ &$S_ p$ & $\Gamma_p$\\ \hline
 1$d_{3/2}$ &  &&&&&&8.3 & 0    \\ 
 2$s_{1/2}$ &  &&&&&&11.5 & 7.7 \\
 1$d_{5/2}$ &  &&&&11.6&0&16.3 & 3.7 \\
 1$p_{1/2}$ &  &&12.1&0&27.5&17.0&33.2 & 21.6 \\
 1$p_{3/2}$ &16.0&0&18.4&3.1$\times$10$^{-6}$&27.5&17.0&33.2& 21.6 \\
 1$s_{1/2}$ &33.9&12.1&41.1&19.0&46.5&21.0&  56.3 & 30.6 \\ \hline
\end{tabular}
\label{tb:sp}
\end{center}
\end{table} 
\end{center}
The effective number $N_{\rm eff}$ is defined as 

\begin{eqnarray}
N_{\rm eff} = \sum_{J M m_s}
\Bigl|\int d^3r 
\chi^{\ast}_f(\mbox{\boldmath $r$}) \xi^{\ast}_{1/2,m_s}
[\phi^{\ast}_{l_K}(\mbox{\boldmath $r$}) \otimes \psi_{j_p}(\mbox{\boldmath $r$})]_{JM}
\chi_i(\mbox{\boldmath $r$})\Bigr|^2.
\end{eqnarray}

\noindent
The proton and the kaon wavefunctions are denoted by $\psi_{j_ p}$ 
and $\phi_{l_ K}$. We adopt the 
harmonic oscillator wavefunction for $\psi_{j_ p}$. The spin wave 
function is denoted by 
$\xi_{1/2,m_s}$,
and we take the spin average with respect to $m_s$, so as to 
take into account the possible spin directions of 
the protons in the target nucleus. The functions $\chi_i$ and $\chi_f$ are the initial and final distorted waves of the 
projectile and ejectile, respectively. We use the eikonal 
approximation and replace $\chi_f$ and 
$\chi_i$ by employing the relation
\begin{eqnarray}
\chi^{\ast}_f(\mbox{\boldmath $r$}) \chi_i(\mbox{\boldmath $r$}) = \exp (i \mbox{\boldmath $q$} \cdot \mbox{\boldmath $r$})D(z, \mbox{\boldmath  $b$}),
\end{eqnarray}
where $\mbox{\boldmath $q$}$ is the momentum transfer between the projectile and 
ejectile, and the distortion factor $D(z,\mbox{\boldmath $b$})$ is defined as 
\begin{eqnarray}
D(z, \mbox{\boldmath $b$}) = \exp \left[
-\frac{1}{2} \sigma_ {KN} \int^{z}_{-\infty}d z^{\prime}
\rho_A (z^{\prime},\mbox{\boldmath $b$})
-\frac{1}{2} \sigma_{pN} \int^{\infty}_{z}d z^{\prime}
 \rho_{A-1} (z^{\prime},\mbox{\boldmath $b$})
\right].
\end{eqnarray}
Here, the kaon-nucleon and proton-nucleon total cross sections 
are denoted by $\sigma_{KN}$ and 
$\sigma_{pN}$. The functions $\rho_A(z,\mbox{\boldmath $b$})$ and $\rho_{A-1}(z,\mbox{\boldmath $b$})$ are 
the density distributions of the target and daughter nuclei 
in the beam direction coordinate $z$ with impact parameter $\mbox{\boldmath $b$}$, 
respectively.

We calculated the kaonic bound state wavefunctions using the optical potentials 
obtained with the chiral unitary model \cite{hirenzaki00} \ 
and the phenomenological fit, \cite{batty97} as 
described in \S\ref{sec:structure}. 
In the chiral unitary model, the depth of the attractive potential is 
only approximately 50 MeV at the center of the nucleus, which is much weaker than 
the phenomenological potential used in Ref.~\citen{batty97}. 
For the case of the phenomenological potential, there exist kaonic 
nuclear bound states with very large binding energies, for example, 100 -- 200 MeV.  For these 
bound states, the ${\bar K}N$ system cannot decay into $\pi\Sigma$, because of the 
threshold, and hence, the widths of these states are expected to be narrower. 
On the other hand, for chiral unitary potential cases, we do not have 
narrow nuclear states, like those in Refs.~\citen{Kishimoto99} and~\citen{Akaishi02},  
because the decay phase space for the ${\bar K}N$ system to the $\pi\Sigma$ channel 
is sufficiently large, 
owing to the smaller binding energies. 
In order to include 'narrowing effects' for the widths due to the 
phase space suppression, we introduce a phase space factor $f^{\rm MFG}$, defined 
in Ref.~\citen{mares04} \ by Mare$\check{\rm s}$, Friedman, and Gal as,

\begin{equation}
f^{\rm MFG}(E)=0.8f^{\rm MFG}_1(E)+0.2f^{\rm MFG}_2(E) ,
\label{eq:mfg}
\end{equation}
\noindent
where $f^{\rm MFG}_1$ and $f^{\rm MFG}_2$ are the phase space factors for 
$\bar{K} N \rightarrow \pi \Sigma$ decay and 
$\bar{K} NN \rightarrow \Sigma  N$, respectively. These factors are defined as 
\begin{equation}
f^{\rm MFG}_1(E)=\frac{M^3_{01}}{M^3_1}\sqrt{\frac{[M^2_1-(m_{\pi}+m_{\Sigma})^2][M^2_1-(m_{\Sigma}-m_{\pi})^2]}
{[M^2_{01}-(m_{\pi}+m_{\Sigma})^2][M^2_{01}-(m_{\Sigma}-m_{\pi})^2]}}\theta(M_1-m_{\pi}-m_{\Sigma}) ,
\label{eq:mfg1}
\end{equation}
and
\begin{equation}
f^{\rm MFG}_2(E)=\frac{M^3_{02}}{M^3_2}\sqrt{\frac{[M^2_2-(m_ N+m_{\Sigma})^2][M^2_2-(m_{\Sigma}-m_ N)^2]}
{[M^2_{02}-(m_ N+m_{\Sigma})^2][M^2_{02}
-(m_{\Sigma}-m_ N)^2]}}\theta(M_2-m_{\Sigma}-m_ N) .
\label{eq:mfg2}
\end{equation}

Here, the branching ratios of mesic decay and non-mesic decay are 
assumed to be 80$\%$ and 20$\%$. The masses are defined as 
$M_{01}=m_{\bar K}+m_{ N}$, 
$M_1=M_{01}+E$, $M_{02}=m_{\bar K}+2m_{ N}$, 
$M_2=M_{02}+E$, and $E$ is the kaon energy defined as 
$E=T_{K}-T_{p}-S_{p}$, using the same kinematical variables as in Eq. (\ref{eq:DE}). 
We multiply the energy independent kaonic widths $\Gamma_K$ by the phase space factor $f^{\rm MFG}$ in order to introduce the energy dependence due to the 
phase space suppression as

\begin{equation}
\Gamma_{K} \rightarrow \Gamma _{K}(E)=\Gamma_{K} \times f^{\rm MFG}(E) .
\label{eq:mfg_G}
\end{equation}

\subsection{Numerical results}
\begin{figure}[htpb]
\epsfxsize=6cm
\centerline{\epsfbox{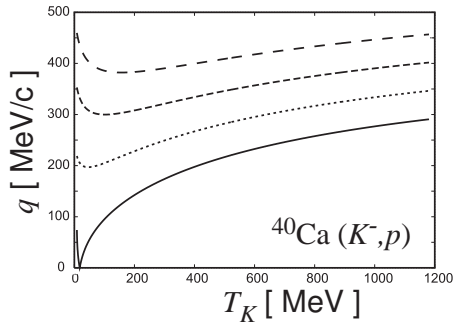}}
\caption{Momentum transfers in the ($K^-,p$) reactions with $^{40}$Ca for targets the 
formation of kaon-$^{39}$K bound systems. The proton separation energy $S_{p}$ is 
fixed at 8.3 MeV, and the kaon binding energies are assumed to be 0 MeV (solid curve), 
50 MeV (dotted curve), 100 MeV (dashed cureve), and 150 MeV (long-dashed curve).}
\label{fig:momentum}
\end{figure}

We present the numerical results for the kaon bound state formation 
spectra in this subsection. 
First, we consider the momentum transfer of the ($K^-,p$) 
reactions as a function of the incident kaon energy, 
which is an important guide to determine suitable incident 
energies in order to obtain a large production rate of the bound states. 
Because we consider both atomic and nuclear kaon states, 
we assume four different binding energies to calculate the 
momentum transfer. We consider the forward reactions and 
the momentum transfer in the laboratory frame, as shown in Fig.~\ref{fig:momentum}. 
As can be seen in the figure, the condition of zero recoil can be 
satisfied only for atomic states with $T_{K} = $10 -- 20 MeV. 
For deeply bound nuclear states, the reaction always requires 
a certain momentum transfer. However, the incident 
energy dependence of the momentum transfer is not strong for 
kaonic nuclear states, as shown in Fig.~\ref{fig:momentum}.

We consider formation of kaonic atoms and kaonic nuclei 
separately, because their properties, and hence, 
the optimal kinematical conditions for their formation are expected to be different. 
We first consider the formation of atomic states. Because the binding energies of 
atomic states are sufficiently small, we can safely ignore the phase space effect for 
the decay widths considered in Eqs. (\ref{eq:mfg}) -- (\ref{eq:mfg_G}). 
For atomic states, the obtained wavefunctions and energy spectra are almost 
identical for the chiral unitary and phenomenological potentials. 
For this reason, we show the results only for chiral unitary potential. 

\begin{figure}[htbp]
\begin{tabular}{cc}
\begin{minipage}{0.5\hsize}
\epsfxsize=6cm
\centerline{\epsfbox{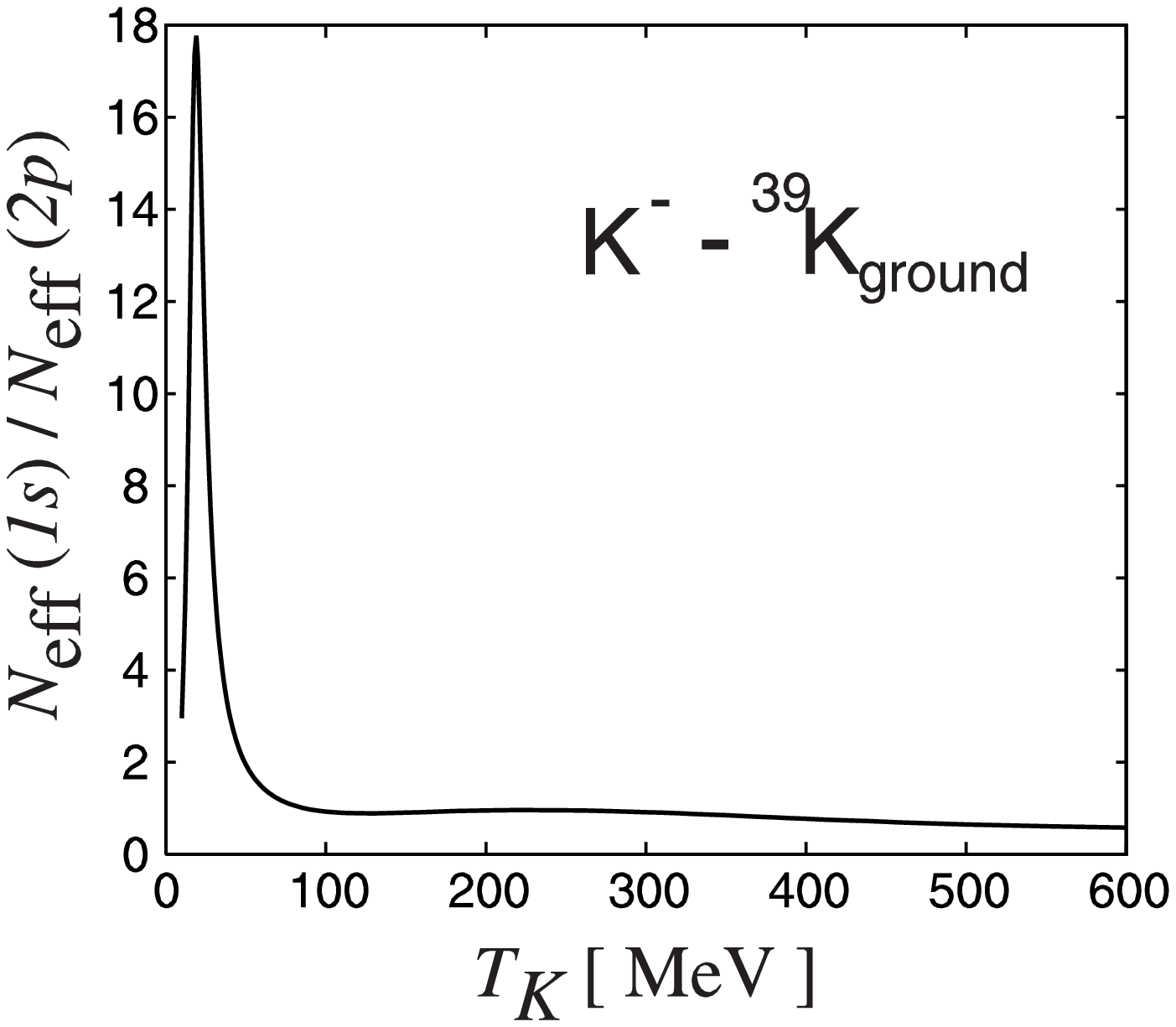}}
\caption{Ratio of the effective numbers of 1$s$ and 2$p$ kaonic atom states formed 
with the [$d_{3/2}^{-1}$] proton-hole state in $^{39}$K plotted as a function of the 
incident kaon energy $T_{K}$.}
\label{fig:1}
\end{minipage}
\begin{minipage}{0.5\hsize}
\epsfxsize=6cm
\centerline{\epsfbox{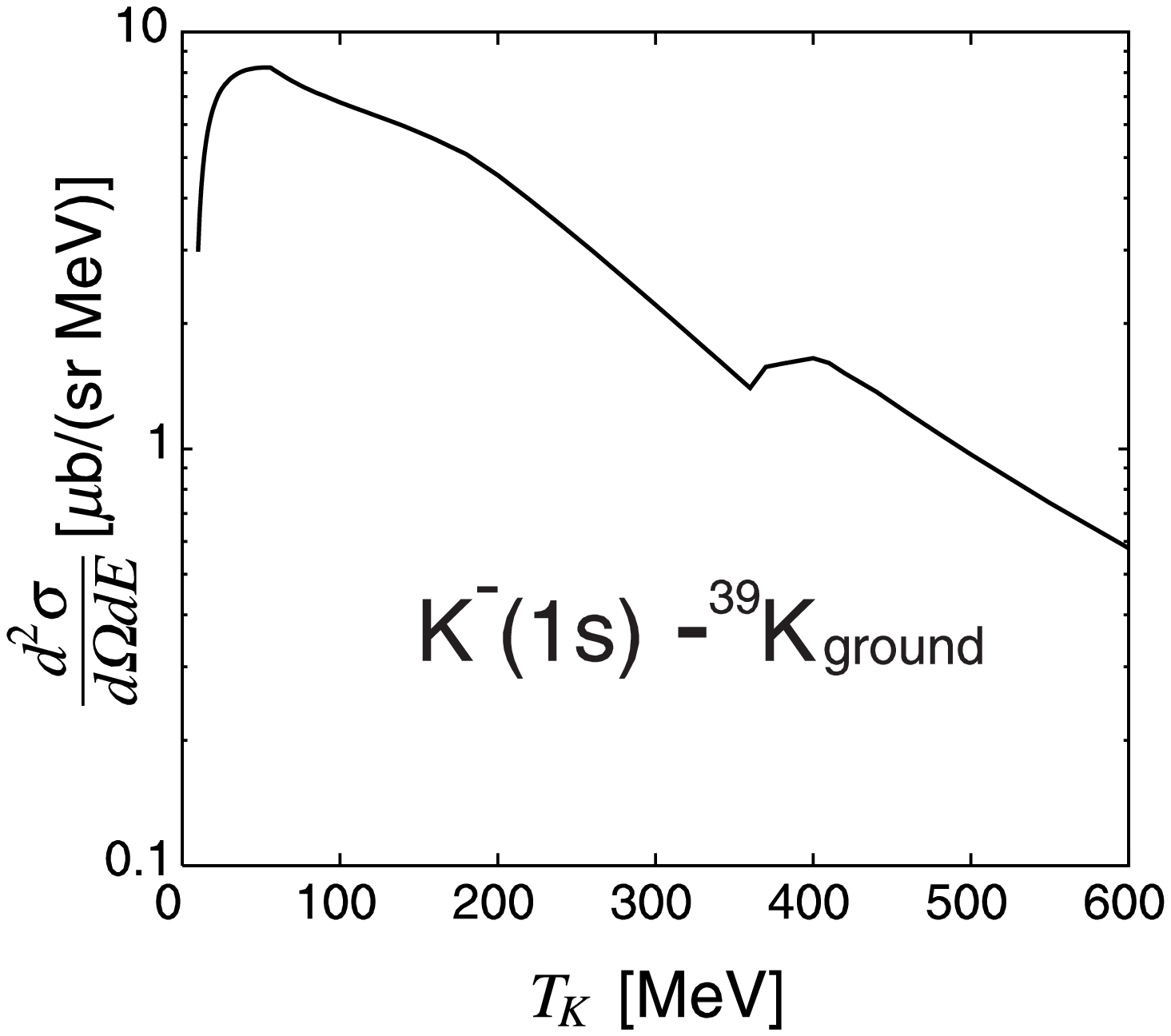}}
\caption{Double differential cross section at the resonance peak energy of the
kaonic atom 1$s$ state formation with the [$d^{-1}_{3/2}$] proton-hole state in $^{39}$K 
at $\theta^{\rm Lab}_{p}$ = 0 [degrees] plotted as a function of the 
incident kaon kinetic energy.}
\label{fig:2}
\end{minipage}
\end{tabular}
\end{figure}

We first study the energy dependence of the atomic 
1$s$ state formation rate in order to 
determine the optimal incident energy for the deepest atomic 1$s$ state 
observation in the ($K^-, p$) reactions. For this purpose, we show in Fig.~\ref{fig:1} \ 
the energy dependence of the 
ratio of the calculated effective numbers of the kaonic atom 1$s$ and 2$p$ states 
coupled to the [$d_{3/2}^{-1}$] proton-hole state in $^{39}$K. 
We find that the contribution of the 
1$s$ state is significantly larger than that of the 2$p$ state at $T_{K}$ = 20 MeV, and 
we therefore hypothesize that the 1$s$ state can be observed clearly without a large 
background due to the 2$p$ kaonic state at this energy, where the 
momentum transfer is reasonably small for atomic state formation. 
Next, we consider the 
energy dependence of the peak height of the 1$s$ kaonic atom state coupled to 
the [$d_{3/2}^{-1}$] proton-hole state in $^{39}$K to determine the 
suitable incident energy to have a large cross section.  As we can see 
in Fig.~\ref{fig:2}, the cross section is maximal somewhere in the range $T_{K}$ = 30 -- 40 
MeV, and we find that the cross section has a local maximum value 
near $T_{K}$ = 400 MeV.  From these observations, we take $T_{K}$ = 20  
and 400 MeV as the incident kaon energies to calculate the energy 
spectrum of the emitted proton.  We also consider $T_{K}$ = 100 MeV as an 
energy between 20 and 400 MeV. We mention here that the eikonal approximation is 
known to be valid only for high energies.  Thus, the results for low energies, i.e. 
$T_{K}$ $\leq$ 100 MeV, should be regarded as rough estimations.  

\begin{figure}[htbp]
\epsfxsize=6cm
\centerline{\epsfbox{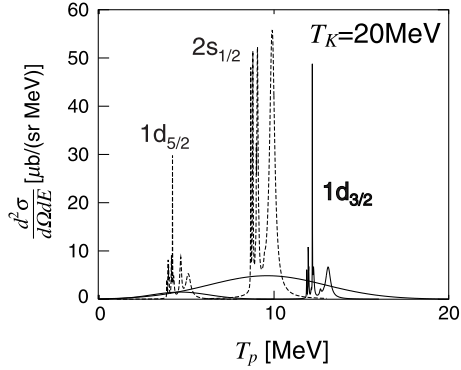}}
\caption{Kaonic atom formation cross sections for each proton-hole state 
in the ($K^-, p$) reactions 
plotted as functions of the emitted proton energy at the incident kaon 
energy $T_{K}$ = 20 MeV and $\theta^{\rm Lab}_{p}$ = 0 [degrees].  
The solid and dashed curves are results with and without 
the widths of the proton-hole states, respectively.}
\label{fig:3}
\end{figure}

In Fig.~\ref{fig:3} we show the calculated spectra for a 
$^{40}$Ca target, including contributions from 
the kaonic atom states up to 4$f$ for [1$d_{3/2}^{-1}$], [1$d_{5/2}^{-1}$] and 
[2$s_{1/2}^{-1}$] proton states in $^{39}$K at $T_{K}$ = 20 MeV.  
The spectra 
without the widths of the proton-hole states $\Gamma_{p}$ 
are represented by the dashed curve.  
We find that the contributions coupled to different 
proton-hole states are localized in different energy regions and are 
well separated from each other.  This feature of the spectra 
is different from that of the pionic atom formation in the ($d,^3$He) reaction,
\cite{hirenzaki91} \ where the contributions from different neutron-hole 
states overlap. 
We find the same features of the spectra for other incident energies.
\cite{okumura00} \ 
The realistic spectra including $\Gamma_{p}$ are plotted by the solid 
lines in the same figure.  As can be seen in the figure, $\Gamma_{p}$ is 
too large to identify each kaonic bound state, 
except for the [$d_{3/2}^{-1}$] hole state, corresponding to the 
ground state of the final daughter nucleus $^{39}$K.  

\begin{figure}[htbp]
\epsfxsize=14cm
\centerline{\epsfbox{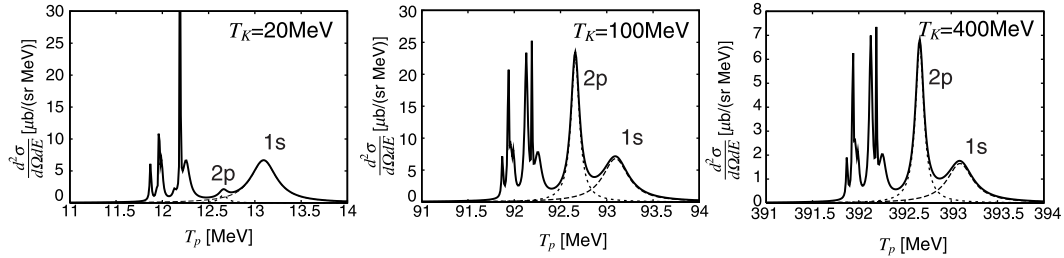}}
\caption{Kaonic atom formation cross sections in $^{40}$Ca($K^-, p$) reactions 
coupled to the [$d_{3/2}^{-1}$] proton-hole state in $^{39}$K 
plotted as functions of the emitted proton energy at 
$\theta^{\rm Lab}_{p}$ = 0 [degrees] at the incident kaon 
energies $T_{K}$ = 20 MeV, 100 MeV and 400 MeV, 
respectively. }
\label{fig:4}
\end{figure}

We show in Fig.~\ref{fig:4} the detailed structure of the kaonic atom formation cross 
section coupled to the ground state of the daughter nucleus $^{39}$K. 
Contributions from deeper proton-hole states provide the smooth 
background of the spectra in this energy region because of their large 
widths.  
We find that the contributions from deeply bound 1$s$ and 2$p$ kaonic atom states 
are well separated. At $T_{K}$ = 20 MeV, the peak due to the 1$s$ state 
is significantly larger than that due to the 2$p$ state, as expected from the 
result in Fig.~\ref{fig:1}.  The peak height of the atomic 1$s$ contribution is approximately 7 
[$\mu b$/($sr$ MeV)] at $T_{K}$ = 20 and 100 MeV.  At $T_{K}$ = 100 MeV, the 2$p$ peak 
is enhanced and is approximately 24 [$\mu b$/($sr$ MeV)].  The overall shapes of the 
cross sections at 100 MeV and 400 MeV are similar, while the 
absolute strength at 400 MeV is approximately 1/3 -- 1/4 of that at 100 MeV. 
We should mention here that the energy resolution of experiments must be 
good enough so as to observe the separate peak stracture in the spectrum 
for the atomic states formation. 

\begin{figure}[htpb]
\epsfxsize=12cm
\centerline{\epsfbox{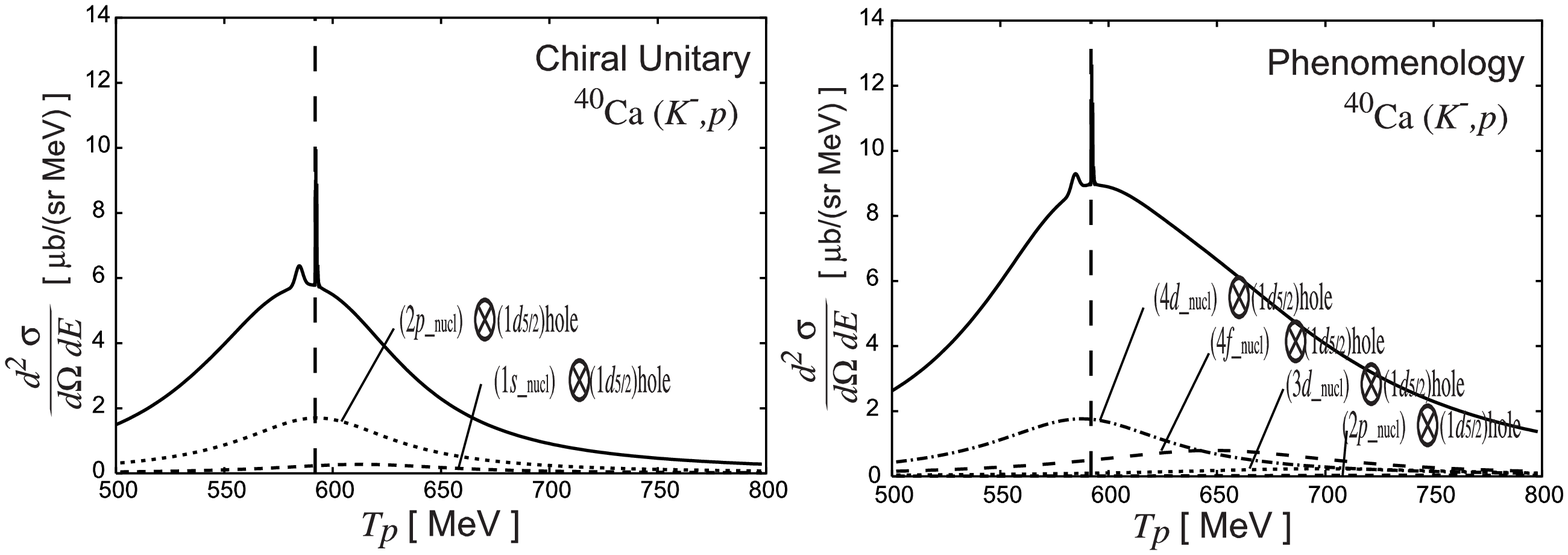}}
\caption{Kaonic nucleus formation cross sections in $^{40}$Ca($K^-,p$) reactions 
plotted as functions of the emitted proton energies at $\theta^{\rm Lab}_{p} = 0$ 
[degrees] and $T_{K} = 600$ MeV for (left) the chiral unitary model and (right) 
the phenomenological $K$-nucleus optical potential. The vertical dashed lines 
indicate the threshold energies, and the sharp peaks around the threshold 
are due to the atomic state formations.}
\label{fig:40Ca}
\end{figure}

We next consider the formation spectra of the kaonic nuclear states in 
the ($K^-,p$) reactions. We choose the incident kaon energy to be 
$T_{K} = 600$ MeV, for which there exist experimental data. \cite{Kishimoto03} 
\ We show in Fig.~\ref{fig:40Ca} the formation spectra of kaonic nuclei 
together with those of kaonic atoms as functions of the emitted proton kinetic 
energies. Here, the widths of the kaonic states are fixed to the values 
listed in Tables \ref{tab:ph} and \ref{tab:chi}. The effects of 
the proton-hole widths $\Gamma_{p}$ are included. We found that 
the spectra do not exhibit any peak-like structure due to the nuclear 
state formation but have only a smooth slope in both the chiral unitary 
and phenomenological optical potential cases. The contributions from the atomic 
state formation appear as two very narrow 
peaks around, the threshold energies 
for both potentials. Each large peak contains several smaller peaks, due to the 
formation of several atomic states, as in the spectrum shown in Fig. \ref{fig:4}.

\begin{figure}[htpb]
\epsfxsize=12cm
\centerline{\epsfbox{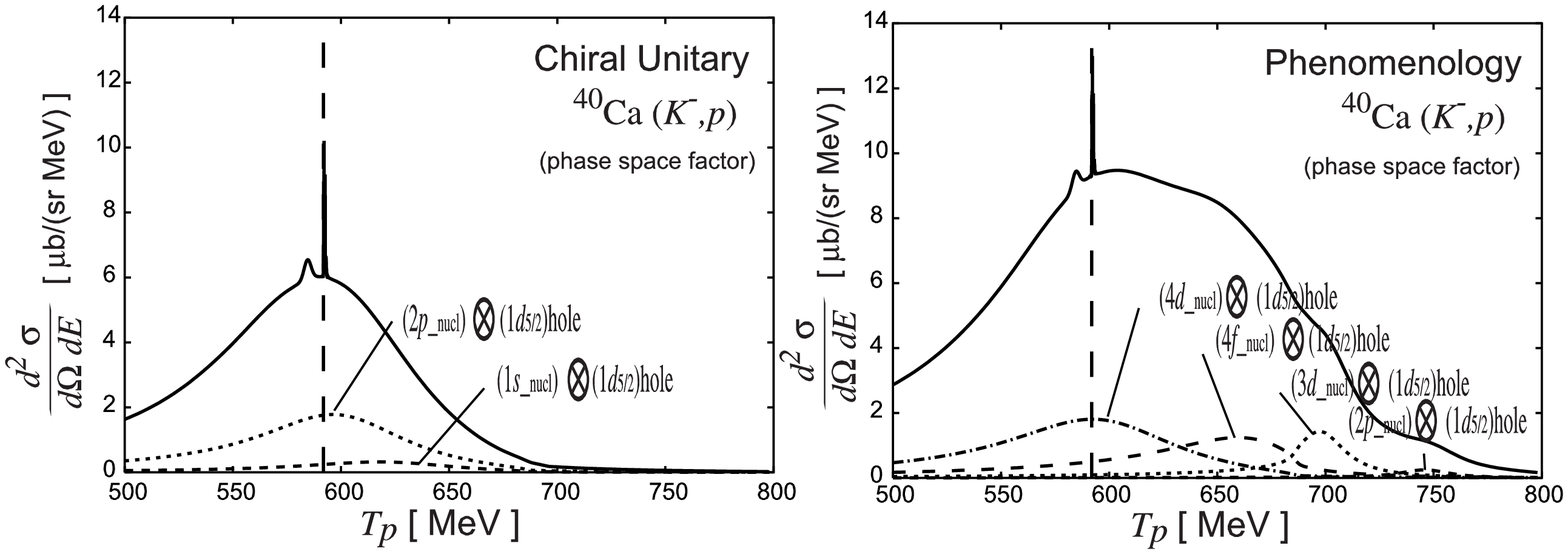}}
\caption{ Kaonic nucleus formation cross sections in $^{40}$Ca($K^-,p$) reactions 
plotted as functions of the emitted proton energies at $\theta^{\rm Lab}_{p} = 0$ 
[degrees] and $T_{K} = 600$ MeV for (left) the chiral unitary model and (right) 
the phenomenological $K$-nucleus optical potential. The vertical dashed lines 
indicate the threshold energies, and the sharp peaks around the threshold 
are due to the atomic state formations. The energy dependent decay 
widths for kaonic states are used. (See the main text in details.)}
\label{fig:39K_phase}
\end{figure}

In order to include the phase space effects on the decay widths for the final kaon 
system, we multiply the widths of kaonic states $\Gamma_K$ by the phase space factor defined in 
Eqs. (\ref{eq:mfg})--(\ref{eq:mfg_G}) 
and calculate the ($K^-,p$) spectra. 
We present the results in Fig. \ref{fig:39K_phase} for both potentials. 
We find that the spectrum for the chiral unitary potential is not 
affected significantly by the phase space effect. However, the ($K^-,p$) 
spectrum shape for the phenomenological potential is distorted by including 
the phase space factor and is expected to possess a bumpy structure, as 
reported in Ref.~\citen{Kishimoto03}.

\begin{figure}[htpb]
\epsfxsize=10cm
\centerline{\epsfbox{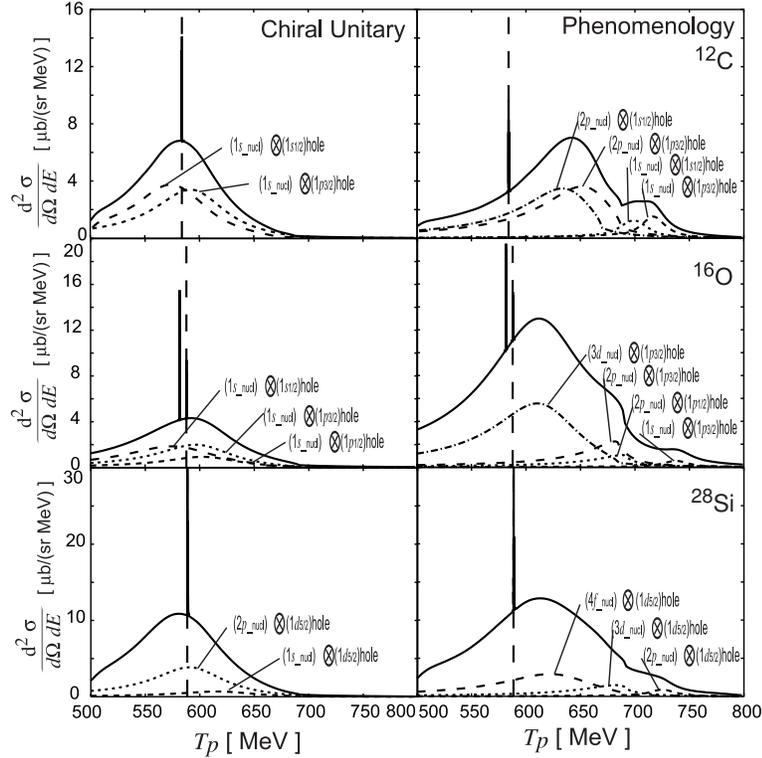}}
\caption{Same as Fig.~\protect\ref{fig:39K_phase}, except that here the target nuclei 
are (top) $^{12}$C, (middle) $^{16}$O, and (bottom) $^{28}$Si.}
\label{fig:spectrum_other}
\end{figure}

We performed systematic calculations for other target nuclei, $^{12}$C, 
$^{16}$O and $^{28}$Si at $T_{K} = 600$ MeV and present the results in 
Fig.~\ref{fig:spectrum_other}. In Ref.~\citen{Kishimoto03}, experimental 
data for a $^{16}$O target are reported and data for $^{12}$C and $^{28}$Si 
targets could be obtained in the future.~\cite{kishimoto_p} 
As shown in Fig.~\ref{fig:spectrum_other}, we have found that 
some bumpy structures in the ($K^-,p$) spectra may appear, due to the formation of 
the kaon nucleus states, especially in the case of the $^{12}$C target, if the 
kaon-nucleus optical potential is as deep as 200 MeV, as reported in Ref.~\citen{batty97}. 
On the other hand, if the depth of the optical potential is as shallow as 
50 MeV, as predicted by the chiral unitary model, the spectrum may not possess any bumpy structures, 
but only exhibit a smooth slope for all targets considered here. 

Finally, we present energy integrated cross sections $\frac{d\sigma}{d\Omega}$ for kaonic nuclear $1s$ state formation for $^{12}$C and $^{28}$Si targets with the results obtained in Refs.~\citen{ciep01} and~\citen{Kishimoto99}. 
We found that our results with the phenomenological optical potential qualitatively 
agree with those in Ref.~\citen{ciep01} and are smaller than those in Ref.~\citen{Kishimoto99}. The present results with the chiral unitary potential are significantly larger than those with the phenomenological potential because of the smaller momentum transfer in the ($K^-,p$) reactions due to the smaller binding energies of kaonic nuclear $1s$ states.
\begin{table}[htpd]
\begin{center}
\caption{Energy integrated cross sections for the formation of kaonic $1s$ nuclear states in units of [$\mu b/sr$]. The results in Refs.~\protect\citen{ciep01} and~\protect\citen{Kishimoto99} are also shown for comparison. Proton hole states are [$1p_{3/2}$]$^{-1}$ and [$1d_{5/2}$]$^{-1}$ for $^{12}$C and $^{28}$Si targets, respectively.}
\begin{tabular}{c|cccc}
\hline
\hline
&\multicolumn{4}{c}{($d\sigma/d\Omega$)$_{(K^-,p)}$ [$\mu b/sr$]}\\
Target nucleus&Chiral Unitary&Phenomenology&Ref.~\protect\citen{ciep01}&Ref.~\protect\citen{Kishimoto99} \\ \hline
$^{12}$C&425&65&47&100--490\\ 
$^{28}$Si&92.6&2.7&6.0&35--180\\
\hline
\end{tabular}
\end{center}
\end{table} 

\section{Conclusion}
\label{sec:conclusion}

We have studied the structure and formation of kaonic atoms and kaonic nuclei in this 
paper. We used two different kaon-nucleus optical potentials, which are obtained 
from the chiral unitary model and a phenomenological fit of existing 
kaonic atom data. We theoretically studied the structure of kaonic atoms and 
kaonic nuclei using these potentials and determined the differences between the obtained level 
schemes of the kaonic nuclear states. 

We also studied the formation cross sections of deeply bound kaonic atoms 
and kaonic nuclei which cannot be observed with standard X-ray spectroscopy.  All the 
atomic states are theoretically predicted to be quasi-stable.  
We 
investigate the ($K^-, p$) reaction theoretically and evaluate the cross 
section of the $^{40}$Ca($K^-, p$) reaction in detail. For deep atomic state formation, the cross sections 
are predicted to be approximately 7 
[$\mu b$/($sr$ MeV)] at $T_{K}$ = 20 and 100 MeV 
and 2 [$\mu b$/($sr$ MeV)] at 400 MeV 
for kaonic atom 1$s$ state formation.  For the atomic 2$p$ state, the cross section is predicted to be approximately 
24 [$\mu b$/($sr$ MeV)] at $T_{K}$ = 100 MeV.  

We also systematically studied the formation cross sections of kaonic nuclear states in 
($K^-,p$) reactions for various targets. 
In order to take into account the phase space suppression effects of the 
decay widths, we introduced a phase space factor to obtain the ($K^-,p$) 
spectra. We found in our theoretically studies that in the ($K^-,p$) reactions, 
a certain bumpy structure due to kaonic nucleus formation can be seen only 
for the case of a deep ($\sim200$ MeV) phenomenological kaon nucleus potential. 
Due to the phase space suppression, the decay widths of kaonic states 
become so narrow that we can see certain bumpy structure in the reaction spectrum, 
which could be seen in experiments. For the case of the chiral unitary potential,
 the binding energies are too small to reduce the decay widths and to see 
the bumpy structure in the spectra of the ($K^-,p$) reactions. However, we should properly include the 
energy dependence of the chiral unitary potential in future studies of kaonic 
nuclear states to evaluate more realistic formation spectra.

In order to obtain more conclusive theoretically results, we need to apply 
Green function methods for states with large widths\cite{morimatsu85} and to consider 
the energy dependence of the optical potential properly. 
Furthermore, we should consider the changes and/or deformations of the nucleus due 
to the existence of the kaon inside and solve the problem in a self-consistent manner 
for kaonic nucleus states. However, we believe that the present theoretical effort 
to evaluate the absolute cross sections for the kaonic bound state formation 
are relevant for determining a suitable method to observe them and helpful for 
developing the physics of kaon-nuclear bound systems and kaon behavior in 
nuclear medium. Further investigations both theoretically 
and experimentally are needed to understand kaon behavior in nuclear medium more 
precisely.

\section*{Acknowledgements}
We acknowledge E. Oset and A. Ramos for stimulating 
discussions on kaon bound systems and the chiral unitary model. 
We would like to thank 
M. Iwasaki and T. Kishimoto for stimulating discussions on the latest experimental data 
of kaonic nucleus formation. We also would like to thank T. Yamazaki,
 Y. Akaishi, and A. Dot$\acute{\rm e}$ for useful discussions on theoretical 
aspects of kaonic nucleus systems. We are grateful to H. Toki and E. Hiyama 
for many suggestions and discussions regarding the kaon-nucleus systems. 
We also thank A. Gal for his careful reading of our preprint and useful comments. 
This work is partly supported by Grants-in-Aid for scientific research of MonbuKagakusho and Japan Society for the Promotion of Science (No. 16540254).

\end{document}